\documentclass{emulateapj}

\shorttitle{UHECR ACCELERATION IN CEN\,A}
\shortauthors{Honda}

\begin{document}

\title{ULTRA-HIGH ENERGY COSMIC-RAY ACCELERATION IN THE
JET OF CENTAURUS\,A}
\author{Mitsuru~Honda}
\affil{Plasma Astrophysics Laboratory, Institute for Global Science,
Mie, Japan}

\begin{abstract}
We evaluate the achievable maximum energy of nuclei diffusively
accelerated by shock wave in the jet of Cen\,A, based on
an updated model involving the stochastic magnetic fields
that are responsible for recent synchrotron X-ray measurements.
For the maximum energy analysis, conceivable energy constraints
from spatiotemporal scales are systematically considered for
the jet-wide including discrete X-ray knots.
We find that in the inner region within $\sim 1\,{\rm arcmin}$ from
galactic core, which includes knots AX and BX, proton and iron nucleus
can be accelerated to $10^{19}-10^{20}$ and $10^{21}~{\rm eV}$
($10-100~{\rm EeV}$ and ${\rm ZeV}$) ranges, respectively.
The upper cutoff energy of the very energetic neutrinos
produced via photopion interaction is also provided.
These are essential for identifying the acceleration site of
the ultra-high energy cosmic ray detected in the
Pierre Auger Observatory, which signifies the arrival from
nearby galaxies including Cen\,A.
\end{abstract}


\section{INTRODUCTION}
To date, the 27 ultra-high energy cosmic ray (UHECR) events of the
energy exceeding $5.7\times 10^{19}~{\rm eV}$ have been detected in
state-of-the-art Auger observatory;
in particular, the anisotropy and significant correlation with the
active galactic nuclei (AGNs) residing within $75~{\rm Mpc}$, namely
GZK horizon \citep{greisen,zatsepin}, were discovered \citep{abraham07}.
Surprisingly, the results indicate that two of these UHECR events
have arrived within $3\degr$ of Centaurus\,A (NGC\,5128), a closest
galaxy \citep[$3.7~{\rm Mpc}$ from us;][]{ferrarese}.
Although the detailed position of the UHECR production site is still
unresolved, the galactic core accompanied by a supermassive
black hole, bipolar jets, giant radio lobes \citep{hardcastle09},
and so on, will be enumerated as the favored candidates.
It is desired that genuine theoretical survey be expanded for clarifying
the particle acceleration mechanism feasible at these sites, in order to
provide the physical interpretation of the observed intriguing results,
which include the recent detection of high-energy gamma rays \citep{aharonian}.

When closely looking at the anatomy of the large-scale jets in the
Cen\,A galaxy \citep[see,][for a review]{israel}, a clue to solve
this challenging problem seems to be in the knotty regions
resolved in a deep X-ray image \citep[e.g.,][]{kraft02}.
In an inner region near the galactic core, the rugged features
are associated with the shocks, which are considered to be formed
where a flowing plasma collides with obstacles \citep{hardcastle07}.
According to the latest theory of plasma kinetic transport,
it is known that such a colliding plasma is likely unstable
for the electromagnetic current filamentation instability,
which generates small-scale magnetic fluctuations with
the order of plasma skin depth (\citealt{medvedev};
\citealt{honda04} and references therein).
In the nonlinear phase, the filaments preferentially coalesce one another,
self-organizing larger scale filaments with stronger magnetic fields.
This reflects a stochastic nature of the inverse energy cascade
of plasma turbulence.
The magnetic fluctuations are non-perturbative, and in the strong
turbulence regime, in that the energy density is comparable to the
thermal energy density of plasma bulk \citep{honda00a,honda00b}.

Indeed, recent high-resolution observations reveal that the jet
of Cen\,A is in part dominated by the filamentary, sometimes
edge-brightened features \citep{hardcastle07}, as inspected in the
well-confirmed jet of a nearby galaxy \citep[Vir\,A;][]{owen}.
Worth noting thing is that inherent inner structure similar to these
has been discovered in many other jets (e.g., Cyg\,A: \citealt{perley};
3C\,353: \citealt{swain}; 3C\,273: \citealt{lobanov};
3C\,438: \citealt{treichel}; Mrk\,501: \citealt{piner}).
Also, it was recently found that a filamentary jet model
naturally provides the comprehensive explanation for the
complicated spectral variability observed in a
TeV blazar object \citep[Mrk\,421;][]{honda08}.
These ingredients strongly encourage us to take the inhomogeneous
magnetic field effects into account for precisely modeling
the Cen\,A jet as a cosmic-ray accelerator.

For a standard, diffusive shock acceleration (DSA) scenario
\citep[e.g.,][]{blandford87}, the strong turbulence
plays an essential role in particle scatterers is expected.
In the stochastic medium, the higher energy heavy particles tend to
freely meander \citep{hh05}, albeit electrons (and positrons) are,
if anything, likely gyro-trapped in the magnetized filaments,
suffering stronger radiative cooling (Section~3.1).
From the fact that in the Cen\,A jet, electron synchrotron emissions
appear to be, in part, already diffusive, it is thus inferred that for nuclei
(say, proton) the conventional approximation of small-angle resonant
scattering will be inadequate for describing the spatial diffusion, and
instead, the three-dimensional rms deflection becomes rather feasible.
Importantly, the latter facilitates the back and forth of
particles across the shock, increasing the efficiency of DSA.
Based on this notion, \citet{hh04} have argued that
a nucleus could be accelerated to the UHE range at a bright jet knot
of Vir\,A, though the arrival from the galaxy ($\sim 4$ times
more distant than the distance to Cen\,A) is not yet signified
\citep[see, e.g.,][for the useful discussions]{stanev}.

Besides, the bound electrons are co-accelerated, to emit the
synchrotron photons, which could serve as a target of
the accelerated protons.
Then, the resonant interaction of the UHE protons with the
target photons could be a major neutrino production process
via the decay channels triggered by photopionization:
$p\gamma\rightarrow \pi^{\pm}X\rightarrow\mu^{\pm}\nu_{\mu}
\rightarrow e^{\pm}\nu_{e}\nu_{\mu}$ \citep{romero}.
The emitted neutrinos propagate on the straight, unlike the
charged particles that suffer, more or less, deflection by
intergalactic magnetic fields \citep[e.g.,][]{vallee}, so that
they play the role of a powerful and complementary messenger of
the UHECR production, in light of the observability at a
forthcoming kilometer neutrino telescope \citep[e.g.,][]{halzen}.
Recently, \citet{cuoco} have proposed the model spectra of high-energy
neutrinos from Cen\,A, but putting the detailed mechanisms
of the particle accelerator into effect has remained unsolved.

In this paper, based on the filament model, we estimate
the maximum possible energies of a proton and iron
diffusively accelerated in the Cen\,A jet, and also,
the energy of the produced neutrinos,
providing the energy-equipartition among the flavors.
The conceivable mechanisms of the energy restriction are taken
into consideration for the jet wide including bright knots.
It is addressed that the energy is limited dominantly by the
shock operation time or particle escape loss, rather than
radiative loss and nucleus--nucleus collision timescales.
As a result, we find that in the inner region, which contains
the X-ray knots BX$n$, proton and iron can be accelerated
beyond the Auger limit.
In particular, the expected highest energy reaches, for iron,
$3~{\rm ZeV}$ and more (around knot BX5); that is to say,
the Cen\,A jet is a "Zevatron" worthy of the candidate
(in addition to the Vir\,A, put forth by \citet{hh04}).

The present analysis virtually provides a substantially extended
version of the previous simple analysis by \citet{romero}, and hence,
a particular attention has been paid to highlight the new points
including (1) the proposal of the improved theoretical model
compatible with updated X-ray measurements (Section~2.1) and
extended arguments on the turbulent magnetic field (Section~2.2),
(2) the non-resonant diffusion scenario involved in the DSA
(Section~3.1), and (3) thorough survey of the temporal (Section~3.2)
and spatial (Section~3.3) limits, in order to figure out the
maximum energies of proton and iron achievable in the
inner region (Section~3.4), and estimate neutrino energy (Section~3.5).
The application of the maximum energy analysis to the
outer region of the large-scale jet is also provided (Section~4).
At last, the discussion on the comparison with the
previous relevant results is expanded (Section~5).

\section{JET WIDE AS A CANDIDATE UHECR FACTORY}

\subsection{\it The Kinematic and Spectral Properties}

Cen\,A is a giant elliptical galaxy (classified as Fanaroff--Riley (FR)
type-I), which contains the bipolar jets ejected from the galactic core.
The brighter jet extends toward the northeast direction beyond
the projection of $\theta_{\rm proj}\simeq 250\arcsec$
\citep[$\theta_{\rm proj}=1\arcsec$ corresponds to
$17~{\rm pc}$;][]{hardcastle07}, with its inclination
with respect to the line of sight (the angle is
referred to as the viewing angle, throughout).
The apparent speed of motion is subluminal, exhibiting
around $0.5c$, where $c$ is the speed of light
\citep{hardcastle03,hardcastle06,horiuchi,brookes}.
Concerning the knots distributing along the jet, a model in which
they are simply compressions in the fluid flow has been ruled out;
and instead, they are considered to be privileged sites
for the in situ particle acceleration
\citep[see,][for a detailed discussion]{hardcastle03}.
Although the velocity distribution inside the knot-like features
is still not revealed, the knots may trace the stationary shocks,
which could be formed as a result of interaction between the jet
fluid and internal obstacles \citep{blandford79a,blandford79b}.
For updated spectral indices around $\alpha_{r}\simeq 0.7$
in observed radio and radio-infrared continuum
\citep{brookes}, we read off the energy spectral index
of energetic electrons of $p=2\alpha_{r}+1=2.4$.
The radiation spectrum extending to the X-ray band is considered
to be of the electron-synchrotron originator, which
contains a key information to understand
the real mechanism of the accelerator that simultaneously
operates for ions, as long as ion abundance is finite
in the jet interior \citep[e.g.,][]{evans,markowitz}.
In this sense, below we deal with somewhat more details
of the X-ray spectrum.

Recent {\it Chandra} observations have successfully resolved the
X-ray knots smaller than the width of the radio-emitting column
with diffuse X-ray emissions \citep{kraft02,hardcastle07}, although
the inner structure of the knots remains unresolved as yet.
When the filamentary morphology inspected in the jet
(introduced in Section~1) retains a self-similar characteristic
\citep{honda08}, the magnification of the spatial scale of
the X-ray knots could repetitiously reveal the composition of
smaller scale compactors and diffuse emitters.
If this insight is correct, the DSA property is incurred
by the filamentary turbulent state \citep{hh04,hh05}.
Considering a similar situation that allows for magnetized
filamentation, \citet{spitkovsky} have verified that the energy
gains truly occur as particles bounce between the upstream and
downstream regions of a collisionless shock, and found
$p=2.4\pm 0.1$, amenable to the aforementioned value.

According to the argument expanded by \citet{hh07}, the
dominant synchrotron emissions (around the spectral break
of $\sim 10^{14}~{\rm Hz}$) are from the major electrons
bound to the magnetized filaments with the
maximum transverse size (i.e., the maximum coherence length).
The electrons in smaller scale filaments down to a
certain characteristic scale $\lambda_{c}$ can be
accelerated to higher energy, because of the weaker
synchrotron loss in the weaker magnetic fields.
In fine filaments smaller than $\lambda_{c}$, electron
acceleration is limited by the escape from the filaments.
\footnote{Note here that this elementary process does {\it not}
mean the escape (in meaning often said) from an entire system
for particle confinement, like a jet column.
The latter is considered in Section~3.3.}
It is pointed out that the free electrons meandering in the
magnetized filaments contribute to engender diffuse
synchrotron photons (in the fundamental process).
Anyhow, it follows that the spatial size scales of filaments are
identified with the local maximum electron energies.
In this context, particularly, it appears that the electron synchrotron
emissions from the major part, larger than $\lambda_{c}$,
dominantly contribute to constitute an extended (typically X-ray)
continuum of spectral energy distribution, instead of sharp cutoff.
Taking into account of the flux density reduction due to the weak magnetic
intensity, we can evaluate the $F_{\nu}$ spectral index above the
break frequency, to give
\begin{equation}
\alpha_{\rm x}=\frac{(5\beta-1)p-(\beta +3)}{2(3\beta +1)},
\label{eq:1}
\end{equation}
where $\beta$ corresponds to the filamentary turbulent spectral index.
Here, we have assumed the Kolmogorov-type fluctuations
superimposing on the local magnetic fields involved in the filaments.
In Equation~(\ref{eq:1}), we find that, for $p=2.4$, the values
of $\beta=5/3$, $2$, and $3$ (suggested by \citet{honda08})
lead to $\alpha_{\rm x}=1.1$, $1.2$, and $1.4$, respectively.
Interestingly, these are accommodated with the measured
values of both $\alpha_{\rm x}=1.00_{-0.15}^{+0.16}$
in between knots\,A2 and B included in the projection
of $\theta_{\rm proj}\leq 60\arcsec$ \citep{kataoka06} and
$\alpha_{\rm x}=1.2\pm 0.2$ in
$60\arcsec<\theta_{\rm proj}<190\arcsec$ \citep{hardcastle07}.

\subsection{\it Turbulent Magnetic Fields}

It is expected that the magnetic field vectors are almost randomly
oriented inside the jet, and as is, the synchrotron radiation
tends to be diffuse with the reduced polarization, with exception
for that from the electrons bound to a pronounced coherent field.
The field strength might be appropriately characterized by
the rms value \citep[denoted simply as $B$;][]{hh05},
although it is hard to directly determine the exact value from
integrating the turbulent spectral intensity over the jet wide.
For simplicity, we here compare it with an equipartition value.
Considering the emitting volume with the size of $\sim 10\arcsec$
\citep{kraft02,hardcastle06} that is larger than the size supposed
by \citet{kataoka05}, we infer that the typical field strength may be
represented by $100~{\rm\mu G}$ in the inner region including bright
knots\,A and B, and less in the outer region \citep{hardcastle06}.
In Table~\ref{tbl:1}, the revised equipartition value,
denoted as $B_{{\rm eq},\delta=1}$, for the simple case
taking no relativistic beaming effect into account is listed.
The energy density levels ($u_{m}$) are thought of as being
larger than (or, at least comparable to) the radiation energy
density ($u_{\rm rad}$) \citep[cf.][]{bai}.

\begin{table}
\begin{center}
\caption{Summary of the Parameter Values\label{tbl:1}}
\begin{tabular}{lccccc}
\tableline\tableline
Region & $L_{\rm proj}({\rm pc})$\tablenotemark{a} &
$R({\rm pc})$\tablenotemark{a} &
$R_{\rm out}({\rm pc})$\tablenotemark{a} &
$B_{{\rm eq},\delta=1}({\rm\mu G})$\tablenotemark{b} &
$\beta$\tablenotemark{c} \\
\tableline
AX1 & $260$ & $352$ & $819$ & $260$ & $5/3$ \\
AX2 & $310$ & $352$ & $819$ & $250$ & $5/3$ \\
AX3 & $430$ & $352$ & $819$ & $290$ & $5/3$ \\
AX4 & $480$ & $352$ & $819$ & $230$ & $5/3$ \\
BX2 & $980$ & $819$ & $1310$ & $160$ & $5/3$ \\
BX5 & $1100$ & $819$ & $1310$ & $290$ & $5/3$ \\
\tableline
\end{tabular}
\tablenotetext{a}{cf. \citet{kraft02}.}
\tablenotetext{b}{Equipartition values excluding the
beaming effects (cf. Section~2.2).}
\tablenotetext{c}{The spectral index of the
filamentary turbulence (cf. Section~2.1).}
\end{center}
\end{table}

In the radial edge of the jet, uncanceled magnetic fields
in the envelope of the clustered filaments cooperate to
apparently constitute a long magneto-tail
extending to the radial direction \citep{hh04}.
The well-pronounced tail would have an ability to trap plasma,
and this stuff is responsible for the observed edge-brightened
features (mentioned in Section~1; \citealt{hardcastle07}).
Relating to this, it is shown in Figure~1 that the jet opening angle
tends to decrease as away from the core; this trend continues up to knot\,G.
This observational fact seems to suggest that such a magnetic field
certainly exerts the net collimation force for jet wide \citep{hh02},
implying that the transverse correlation length of the magnetized
filamentary turbulence reaches the radial size scale of the jet
(beyond the aforementioned size scale of compact X-ray knots).
Within the present framework, the various size scale of fragmented
pieces inside the jet is considered to reflect the coherence length
of the turbulence (as compatible with the argument given in
Section~2.1), although the real situation might be more complicated
\citep{blandford79a,blandford79b,hardcastle03}.

\begin{figure}
\centerline{\includegraphics*[bb=50.0 50.0 685.0 550.0,
width=\columnwidth]{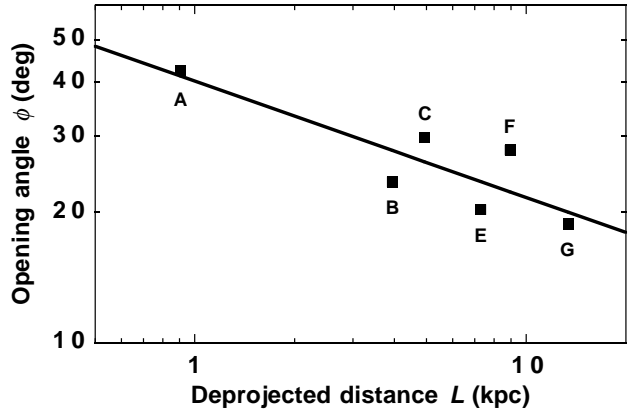}}
\caption{Opening angle of the Cen\,A jet (in the regions
including the X-ray knots; labeled) $\phi$ vs.
the deprojected distance from galactic core $L$.
Here, the definition of $\phi=2\tan^{-1}(R\sin\theta/L_{\rm proj})$
has been used, where
$R$ and $L_{\rm proj}$ are the radius of jet and projected distance
from the core, respectively (cf. Table~\ref{tbl:3}), and $\theta$
is the viewing angle, assumed to be $15\degr$ \citep{hardcastle03}.
The solid line indicates the interpolation among the points.\\
}
\end{figure}

Founded on this simple model, in the following Section~3
we attempt to expand rigorous arguments on the DSA of particles
in the inner region ($\theta_{\rm proj}\leq 60\arcsec$),
in which the knots A and B have been, at least in part, associated
with shock acceleration sites \citep{hardcastle07}.
As the outer jet of $\theta_{\rm proj}>60\arcsec$ concerns,
the X-ray emission is largely diffuse, and the compact knots
embedded in the diffuse emission are much smaller than the
radial sizes of the jet.
Although there is no evidence for shocks (other than at knots)
anywhere in the outer jet of $\theta_{\rm proj}>60\arcsec$ at the moment,
it seems that the appearance of such inner structure as well as the
measured X-ray spectrum can be explained by the present model
including the inhomogeneity, according to the discussions given above.
It is also noted that in the outer region, the presence of
stochastic magnetic fields has been suggested
(\citealt{mao}, within the framework of \citet{fleishman}).
At this juncture, we provide, in the subsequent Section~4,
a preliminary argument parallel to that expanded in Section~3,
considering an application of the model to the issue of
particle acceleration in the outer region
($60\arcsec<\theta_{\rm proj}\leq 200\arcsec$)
including knots C, E, F, and G, where magnetic self-collimation
of the jet seems to be viable (Figure~1).

\section{PARTICLE ACCELERATION IN THE INNER JET REGION}

\subsection{\it Acceleration Timescale}

In the filamentary medium, the diffusion property of
nuclei is quite different from that of electrons.
For instructive manner, let us begin with considering the cold limit,
in which any species of charged particles are gyro-trapped
in the (local) mean magnetic field permeating through a filament.
For the DSA in this regime, the particles will suffer conventional
gyro-resonant diffusion, going back and forth across the shock
discontinuity to gain their energies.
Then, the mean acceleration time is roughly proportional
to $\sim r_{g}/c$, where $r_{g}$ is the gyroradius
(smaller than the radial size of the filament).
In the ordinary case in which radiative mechanism spontaneously impedes
particle energization, the kinetic energies disperse among the
particle species, such that the ratio of the resulting gyro-radii
among proton, arbitrary ion, and electron would indicate
\begin{equation}
r_{g,p}:r_{g,i}:r_{g,e}=1:\frac{1}{\sqrt{Z}}\left(\frac{A}{Z}\right)^{2}:
\left(\frac{m_{e}}{m_{p}}\right)^{2},
\label{eq:2}
\end{equation}
\noindent
respectively, where $Z$ and $A$ are the charge and mass number
of nuclei, and $m_{e,p}$ is the electron and proton mass, respectively.
That is, we read off $r_{g,p}\sim r_{g,i}$, and
$r_{g,p}/r_{g,e}\sim O(10^{6})$.
The result indicates that the proton (ion) escape from magnetized
filaments, which is associated with the bound-free transition
\citep{hh05}, is easier than the electron escape.

Unlike the major electrons quasi-secularly bound to the filaments
\citep[Section~2.1;][]{hh07,honda08}, energetic ions are likely to
freely meander in the forest of filaments, undergoing the
three-dimensional rms deflection.
The latter is in the regime of the off-resonant scattering
diffusion, which still makes possible to go back and forth
across the shock, to additionally accelerate the transitionally
injected ions to a higher energy range \citep{hh04,hh05}.
It is mentioned that this original idea could be reconciled with
the recent simulation results for the similar plasma configuration
\citep[Section~2.1;][]{spitkovsky}.

When considering the ideal Fermi type-I mechanism,
the energy spectral index of accelerated ions
ought to be identical to that of co-accelerated electrons
\citep[see, e.g.,][for details]{schlickeiser},
and thus, denoted commonly as $p$.
The mean acceleration time for arbitrary nuclei
is found to be given by
\begin{equation}
t_{\rm acc}\simeq\frac{\sqrt{6}\pi a_{1}(\beta,p)}{8}
\frac{c}{R}\left(\frac{E}{ZeBU}\right)^{2},
\label{eq:3}
\end{equation}
\noindent
where $a_{1}$ is the dimensionless factor
\begin{equation}
a_{1}(\beta,p)=\frac{\beta(2p+1)}{(\beta-1)(p^{2}-1)},
\label{eq:4}
\end{equation}
\noindent
which is of order unity, and $e$, $E$, $R$, and $U$ are the
elementary charge, particle energy, (transverse) correlation length
of the filamentary turbulence, and flow speed upstream viewed
in the shock rest frame, respectively \citep{hh05}.
Recalling the arguments given in Section~2.2, $R$ is reasonably compared
with the radius of the jet, rather than that of compact knots.
It is noted that Equation~(\ref{eq:3}) is valid for the unbound
particles, which meander in the magnetized filaments, though
still confined in the system with the size $\sim R$.
Hereafter, we set $p=2.4$, along with the arguments in Section~2.1.
In this Section~3, we pay attention to the inner regions
that wrap the well-resolved small X-ray knots
AX1, AX2, AX3, AX4, BX2, and BX5 \citep{kraft02,kataoka05}.
For convenience, the parameter values (referred for the
analysis in Section~3) are summarized in Table~\ref{tbl:1}.
Equation~(\ref{eq:3}) is used to balance the various
loss timescales evaluated below, to solve for $E$.

\subsection{\it Temporal Limits}

\subsubsection{\it Shock-Accelerator Operation Time}

The dynamical timescales concerning the energy limit of
accelerated particles include the operation time of shock
accelerator and adiabatic expansion loss time.
It is hard to precisely evaluate the former, because of the
difficulty of the identification of shock structure and $U$.
In an ideal case in which the shock is stationary as sustained
by an exhaustless incoming flow, one can ignore this kind of
restriction, so that the maximum energy analysis can be
simplified (the related discussion is given later).
Below, we consider, for heuristic, the generic situation
in which the shock accelerator moves along the jet.
For simplicity, it may be reasonable to compare
$U$ to the moving speed (with respect to the core).
The shock accelerator operation time can then be estimated as
$t_{\rm sh}\sim L/U$, and the adiabatic expansion loss time,
$t_{\rm ad}\sim 3L/(2U_{r})$ \citep[e.g.,][]{muecke},
where $L$ and $U_{r}$ are the length scale of the
jet and the speed of radial expansion, respectively.
For $U>U_{r}$ compatible with the self-collimating
characteristic of jet flow (Section~2.2), we have the
relation of $t_{\rm sh}<t_{\rm ad}$, which suggests that
$t_{\rm sh}$ preferentially limits the particle acceleration.

When taking into account of the relativistic effects, the scaling
of $t_{\rm sh}$ can be explicitly written as follows:
\begin{equation}
t_{\rm sh}=2.1\times 10^{10}\frac{1}{\Gamma(U)}
\frac{L}{1~{\rm kpc}}\frac{0.5c}{U}~{\rm s}.
\label{eq:5}
\end{equation}
\noindent
Here, $\Gamma=[1-(U/c)^{2}]^{-1/2}$,
and $L$ is the deprojected distance of the concerned knotty regions
from the core, that is, $L_{\rm proj}/\sin\theta$, where
$L_{\rm proj}$ and $\theta$ are the projected distance
and viewing angle, respectively (Section~2.1).
According to the discussions given above, we set
$U=U_{\rm app}/[\sin\theta+(U_{\rm app}/c)\cos\theta]$,
where $U_{\rm app}$ stands for the apparent speed of motion.
In Figure~2, we plot $t_{\rm sh}$ as a function of $\theta$,
for the cases of $U_{\rm app}=0.1c$ (top) and $0.5c$ (bottom).
It is noted that the enhancement of $t_{\rm sh}$ for the marginally
smaller $\theta$ is due to the dominant increase of
$L$ against the increase of $U$ and $\Gamma$ in the right-hand side
of Equation~(\ref{eq:5}).
For a plausible value of $\theta=15\degr$
\citep[shaded lines;][]{hardcastle03},
one can find $t_{\rm sh}\sim 10^{11}-10^{12}~{\rm s}$.
For example, for a possible parameter set of $U_{\rm app}=0.1c$ and
$\theta=15\degr$, we figure out $t_{\rm sh}=3.5\times 10^{11}~{\rm s}$
and $1.5\times 10^{12}~{\rm s}$ in the regions with
knots AX1 and BX5, respectively.

\begin{figure}
\centerline{\includegraphics*[bb=45.0 32.5 565.0 730.0,
width=\columnwidth]{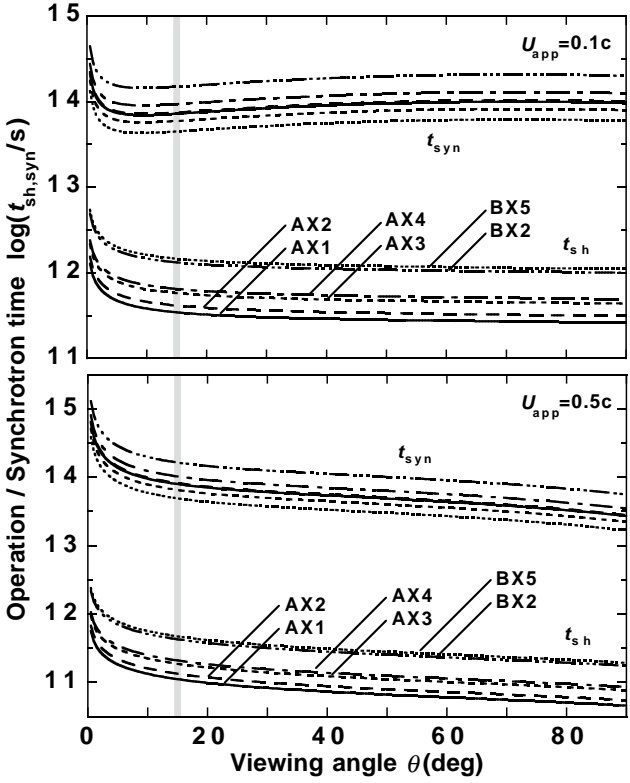}}
\caption{Operation timescales of shock accelerator
$t_{\rm sh}$ for the apparent speed of $U_{\rm app}=0.1c$
(top) and $0.5c$ (bottom) vs. $\theta$, in the
regions around the six inner bright X-ray knots (included in the
projection of $\theta_{\rm proj}\leq 60\arcsec$; labeled).
The synchrotron cooling timescales $t_{\rm syn}$ are also indicated
(the explanatory notes correspond to those for $t_{\rm sh}$).\\
}
\end{figure}

\subsubsection{\it Radiative Losses}

The radiative losses of the charged particles meandering
in the filamentary medium compete with the energization.
We anticipate that the emission is of diffuse \citep{fleishman},
and then, the cooling timescale that can be expressed as
$t_{\rm syn}=36\pi^{2}{\tilde\tau}(\beta)(A/Z)^{4}
[m_{p}^{4}c^{7}/(c^{4}B^{2}E)]$,
where ${\tilde\tau}(\beta)=(6\pi)^{-2}
[\beta(\beta+2)^{2}(\beta+3)]/[2^{\beta}(\beta^{2}+7\beta+8)]$
\citep{hh07}.
For the maximum energy analysis, the temporal balance of
$t_{\rm acc}=t_{\rm syn}$ gives the solution of $E$, and
putting this into the $t_{\rm syn}$-expression denoted above
is found to provide the following practical scaling
\begin{eqnarray}
t_{\rm syn}&=&8.4\times 10^{14}a_{1}(\beta,p)^{1/3}
\left[\frac{{\tilde\tau}(\beta)}{10^{-3}}\right]^{2/3}
\frac{1}{Z^{2/3}}\left(\frac{A}{2Z}\right)^{8/3} \nonumber \\
&&\times\left(\frac{100~{\rm\mu G}}{B}\right)^{2}
\left(\frac{100~{\rm pc}}{R}\right)^{1/3}
\left(\frac{0.5c}{U}\right)^{2/3}~{\rm s}.
\label{eq:6}
\end{eqnarray}
\noindent
Note that setting to $(\beta,p)=(5/3,2.4)$, which reflects
$(\alpha_{r},\alpha_{\rm x})=(0.7,1)$ (cf. Section~2.1), gives
$a_{1}=3.0$ and ${\tilde\tau}=2.4\times 10^{-3}$.
In Equation~(\ref{eq:6}), $B$ is the quantity in the shock rest frame,
and hence, for the moving shock case, set to
$B_{{\rm eq},\delta=1}\delta^{-5/7}$ \citep{kataoka05},
where $\delta=\Gamma^{-1}\left[1-(U/c)\cos\theta\right]^{-1}(>1)$
is the beaming factor, and $U=U(U_{\rm app},\theta)$ (Section~3.2.1).

Relating to this loss process, here it might be useful to remark
the collision with photons, which also limits the particle energization.
This actually takes place, as far as a finite $u_{\rm rad}$ level is
sustained by the synchrotron radiation from co-accelerated electrons,
galactic emissions, not to mention cosmic background lights.
The timescale of the $p\gamma$ interaction can be estimated as
$t_{p\gamma}\sim (\chi{\tilde\tau})^{-1}(u_{m}/u_{\rm rad})t_{\rm syn}$,
where $\chi$ is a dimensionless factor almost independent
of the source parameters \citep{biermann,romero}.
By invoking $\chi\approx 200$, thereby $\chi{\tilde\tau}\lesssim O(1)$,
we read off $t_{p\gamma}\gtrsim (u_{m}/u_{\rm rad})t_{\rm syn}$,
so that the inequality of $u_{\rm rad}\leq u_{m}$, which is
expected in the concerned jet-knot environments (Section~2.2),
results in $t_{p\gamma}\geq t_{\rm syn}$.
It thus appears that for the maximum energy analysis, it is sufficient
to balance $t_{\rm acc}=t_{\rm syn}$, as carried out above.

In Figure~2 for $U_{\rm app}=0.1c$ ({\it top}) and $0.5c$
({\it bottom}), we show the $t_{\rm syn}$ levels
for protons, i.e., Equation~(\ref{eq:6}) for $(A,Z)=(1,1)$,
using the parameter values listed in Table~\ref{tbl:1}.
For smaller $\theta$, $t_{\rm syn}$ sharply increases,
since $\delta$ increases (i.e., $B$ decreases).
As $\theta$ increases, both $\delta$ and $U$ decrease,
so that their competition influences upon whether $t_{\rm syn}$
decreases or not in the larger $\theta$-region.
For the concerned regions, the timescale ranges from
$t_{\rm syn}=4.5\times 10^{13}~{\rm s}$ (BX5) to
$1.5\times 10^{14}~{\rm s}$ (BX2) for $\theta=15\degr$.
It is found that the $t_{\rm syn}$ level is order of magnitude
higher than $t_{\rm sh}$ over the whole $\theta$-range.
As would be expected, the characteristic of the timescales well
separated markedly simplifies the maximum energy analysis.

\subsubsection{\it Collisional Loss}

The collision of the cosmic-ray (test) particles with
target particles also degrades the acceleration.
In what follows, we clarify the condition for which the
collisional loss can be neglected.
The timescale of $pp$ interaction may be simply
given by $t_{\rm col}=(\sigma_{pp}cn)^{-1}$, where
$\sigma_{pp}\approx 4\times10^{-26}~{\rm cm^{-2}}$ and $n$ are the
cross section and number density of target protons, respectively.
In conjunction with the issue of neutrino production, the
$pp$ collision involves the possible reaction channel
of $p+p\rightarrow 2d+e^{+}+\nu_{e}+\gamma$.
Here we derive, for convenience, the critical density
of target protons, denoted as $n_{\rm cr}$, above which
$t_{\rm col}<{\rm min}(t_{\rm sh},t_{\rm syn})\sim t_{\rm sh}$ sets in.

In Figure~3, we plot $n_{\rm cr}$ against $\theta$.
The $n_{\rm cr}(\theta)$ depicted here is for the case of smaller
$U_{\rm app}=0.1c$, since the value is lower (thereby,
more crucial) than that for the larger $U_{\rm app}$ case.
For the inclination of $\theta=15\degr$, the critical density
ranges from $5.7\times 10^{2}~{\rm cm^{-3}}$ (BX5)
to $n_{\rm cr}=2.4\times 10^{3}~{\rm cm^{-3}}$ (AX1).
It turns out that, if $n<10^{2}~{\rm cm^{-3}}$ is satisfied,
$t_{\rm col}$ cannot be, in the whole $\theta$-range,
the shortest timescale among the energy loss processes.
For example, the expected value of the order of
$n\sim 10^{-1}~{\rm cm^{-3}}$ \citep{meisenheimer} leads to
$t_{\rm col}\gg {\rm min}(t_{\rm sh},t_{\rm syn})$.
That is to say, one can almost ignore the
collisional effects on the energy loss.
It follows that the $pp$ neutrino production channel is less
outweigh than the $p\gamma$ one that is considered in Section~3.5.

\begin{figure}
\centerline{\includegraphics*[bb=70.0 32.5 685.0 550.0,
width=\columnwidth]{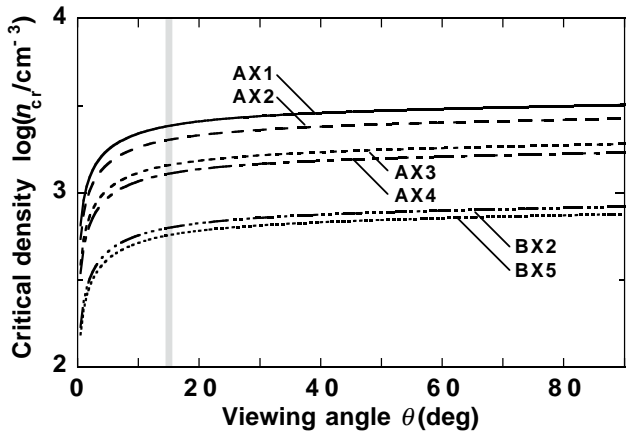}}
\caption{Critical density $n_{\rm cr}$ in each regions
(labelled), above which the proton--proton collisional timescale
$t_{\rm col}$ is shorter than $t_{\rm sh}$ (for $U_{\rm app}=0.1c$).
For $\theta=15\degr$ (shaded line), it is evident that
the collisional loss of an accelerated proton is less dominant,
if $n<n_{\rm cr}=10^{2}-10^{3}~{\rm cm^{-3}}$.\\
}
\end{figure}

Summarizing the discussions in Section~3.2, we can claim
that the temporal balance equation can be reduced as
$t_{\rm acc}={\rm min}(t_{\rm sh},t_{\rm syn},t_{\rm col})\sim t_{\rm sh}$,
when considering the finite dynamical timescale of a moving shock.
The solution for $E$ is denoted as $E_{t}$, which indicates
the maximum possible energy of an accelerated particle,
determined by the temporal limits.
This value should be compared with the energy restricted
from spatial scale; the related issue is discussed below.

\subsection{\it Spatial Limit}

In the present context, when the mean free path of accelerated
particles, transverse to the filaments ($\lambda_{\perp}$),
reaches the order of the radial size of the jet ($\sim R$), the
particles would start escaping from the concerned system.
This is just a subject to the spatial limit of
particle acceleration.
The balance equation $\lambda_{\perp}(E)=R$, contrast to $r_{g}=R$
for the conventional gyro-resonant DSA, appropriately provides the
solution for $E$ (denoted as $E_{s}$), which indicates the maximum
possible energy determined by the spatial limit.
Making use of the expression of $\lambda_{\perp}$ that
can be derived from the quasi-linear kinetic equation
\citep[see][for the details]{hh05}, we obtain
\begin{equation}
E_{s}=4.3\times 10^{18}a_{2}(\beta,p)Z
\frac{B}{100~{\rm\mu G}}
\frac{R}{100~{\rm pc}}~{\rm eV},
\label{eq:7}
\end{equation}
\noindent
where $a_{2}$ is the dimensionless factor
\begin{equation}
a_{2}(\beta,p)=\left[
\frac{(\beta -1)(p+2)(p+4)}{\beta}
\right]^{1/2}.
\label{eq:8}
\end{equation}
\noindent
Note that for $(\beta,p)=(5/3,2.4)$, Equation~(\ref{eq:8})
gives $a_{2}=3.4$.

If the outer ridge with the size exceeding $R$ is pervaded by
a large-scale magnetic tail (Section~2.2), we could speculatively
regard such a size as effective confinement radius.
When one allows for setting to , e.g., the value around
$\sim R_{\rm out}(>R)$ listed in Table~1 \citep{kraft02},
the $E_{s}$ value (given in Equation~(\ref{eq:7}))
is found to be raised by a factor of $\sim\sqrt{R_{\rm out}/R}$.
However, this factor is of the order unity, and it can be
confirmed that the effect does not severely affect the
conclusions for the simple case excluding them.
In this aspect, here we show the analysis, for which
Equation~(\ref{eq:7}) is referred in the rather conservative
manner, to provide the theoretical upper limit of particle energy
which is of the form that can be made a direct comparison with $E_{t}$.

\subsection{\it Highest Energies of Proton and Iron}

For the concerned jet regions each, we evaluate the
achievable maximum energy that is defined by
$E_{\rm max}={\rm min}(E_{s},E_{t})$, where $E_{s}$ is
given by Equation~(\ref{eq:7}), and $E_{t}$ can be
properly derived from $t_{\rm acc}=t_{\rm sh}$
as argued in Section~3.2, to have the scaling of
\begin{eqnarray}
E_{t}=&&2.1\times 10^{19}\frac{1}{a_{1}(\beta,p)^{1/2}}Z
\frac{B}{100~{\rm\mu G}} \nonumber \\
&\times&
\left(\frac{L}{1~{\rm kpc}}\right)^{1/2}
\left(\frac{R}{100~{\rm pc}}\right)^{1/2}
\left(\frac{U}{0.5c}\right)^{1/2}{\rm eV}.
\label{eq:9}
\end{eqnarray}
\noindent
Hereafter, the solutions $E_{\rm max}$ for proton $(A,Z)=(1,1)$
and iron $(56,26)$ are, for convenience, simply
denoted as $E_{p,m}$ and $E_{i,m}$, respectively.

In Figure~4, for $Z=1$ we plot $E_{s}$ (Equation~(\ref{eq:7}))
and $E_{t}$ (Equation~(\ref{eq:9})), using the
parameter values listed in Table~1.
One can see that $E_{p,m}={\rm min}(E_{s},E_{t})=E_{t}$
is ensured in the major range of $\theta>5\degr$,
while in the marginally smaller $\theta$,
the spatial limit regime with $E_{p,m}=E_{s}$ can appear.
Particularly, for a standard $\theta=15\degr$,
it is sure that $E_{p,m}=E_{t}$ is valid for both
$U_{\rm app}=0.1c$ and $0.5c$ in all concerned regions.
This is one of the most noticeable results, which is different from
the previous one derived from the analysis for a narrower jet
with $R<100~{\rm pc}$ \citep[Vir\,A;][]{hh04}, where
${\rm min}(E_{s},E_{t})=E_{s}$ appears in a possible $U$ range.
For practical reasons, the resulting values of $E_{p,m}$
for $U_{\rm app}=0.1c$ are listed in Table~\ref{tbl:2}.
It is noted that the $U_{\rm app}$-dependence of $E_{p,m}$
is weak around $\theta=15\degr$, as seen in Figure~4.
We can conclude that the $E_{p,m}$ value more likely
reaches as high as $\sim 10^{20}~{\rm eV}$ in the inner jet
region ($\theta\leq 60\arcsec$) including knots BX$n$.
Also, the values of $E_{i,m}$ have been derived
in parallel to $E_{p,m}$, and added in Table~\ref{tbl:2}.
It is found that the inner regions can be iron
Zevatron; particularly, in the region with knot BX5,
$E_{i,m}$ reaches $3.2\times 10^{21}~{\rm eV}$.

\begin{figure*}[t]
\centerline{\includegraphics*[width=15.5cm]{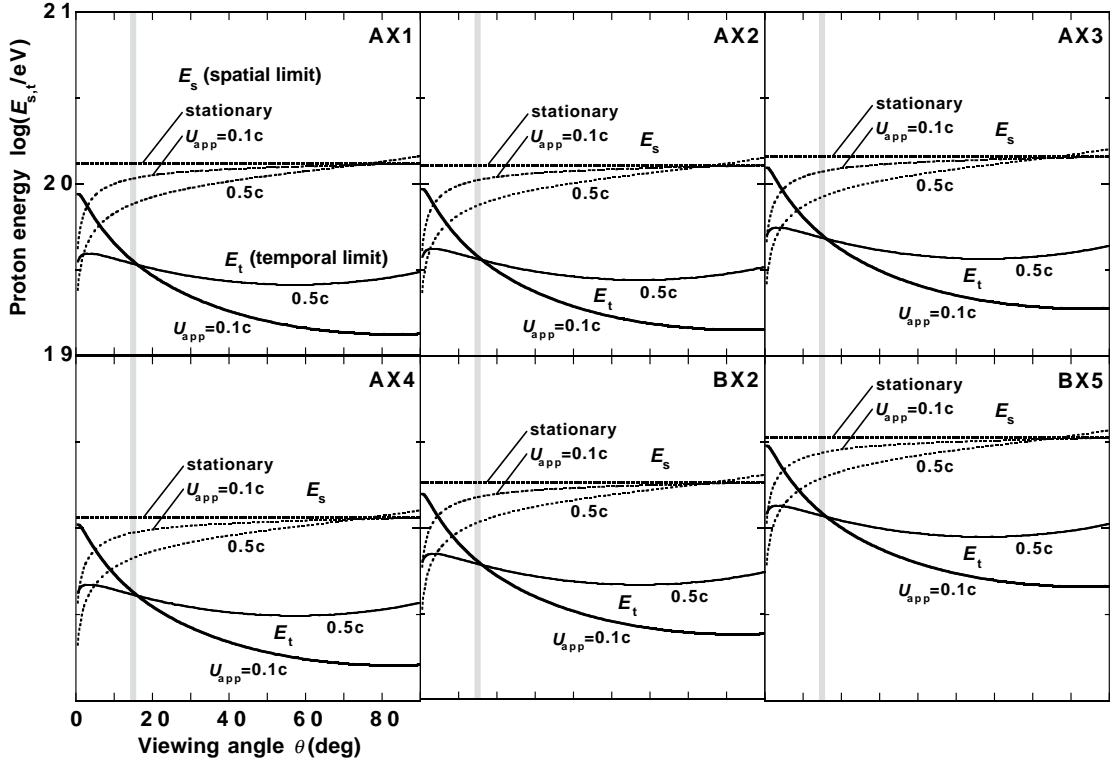}}
\vspace{-.2cm}
\caption{Proton energies $E_{s}$ and $E_{t}$
determined by the spatial and temporal limits, respectively,
in each regions (labeled).
The axes are common in the panels.
Note that $E_{t}(\theta)$ is the solution for
${\rm min}(t_{\rm sh},t_{\rm syn},t_{\rm col})=t_{\rm sh}$,
given $U_{\rm app}$ (labeled).
The achievable maximum energy is given by
$E_{\rm max}={\rm min}(E_{s},E_{t})$; for both
$U_{\rm app}=0.1c$ and $0.5c$, $E_{\rm max}=E_{t}$ is
satisfied at $\theta=15\degr$ (shaded lines), and
for the stationary shock case, $E_{\rm max}=E_{s}$.
The $E_{\rm max}$ values (denoted as $E_{m,p}$ for proton)
are summarized in Table~\ref{tbl:2}.\\
}
\end{figure*}

\begin{table}
\begin{center}
\caption{The Maximum Energies of Proton, Iron,
and Neutrino\label{tbl:2}}
\begin{tabular}{lccc}
\tableline\tableline
Region & $E_{p,m}({\rm EeV})$ & $E_{i,m}({\rm ZeV})$ &
$E_{\nu,m}({\rm EeV})$\\
\tableline
AX1 & $35\,(130)$ & $0.92\,(3.4)$ & $1.8\,(6.5)$\\
AX2 & $38\,(130)$ & $0.99\,(3.3)$ & $1.9\,(6.4)$\\
AX3 & $50\,(140)$ & $1.3\,(3.8)$ & $2.5\,(7.2)$\\
AX4 & $43\,(120)$ & $1.1\,(3.0)$ & $2.1\,(5.8)$\\
BX2 & $64\,(180)$ & $1.7\,(4.8)$ & $3.2\,(9.2)$\\
BX5 & $120\,(340)$ & $3.2\,(8.7)$ & $6.1\,(17)$\\
\tableline
\end{tabular}
\tablecomments{Values in the round brackets indicate
those for the stationary shock case.}
\end{center}
\end{table}

Here, we provide the discussion on the special case
in which the shock is stationary \citep{hardcastle03}.
The operation of DSA continues until the incoming flow into
the shock breaks off; otherwise, the accelerator secularly operates.
In the former case, it is hard to determine $t_{\rm sh}$, because of
the complexity of the spatiotemporal dynamical evolution of the jet.
For the present maximum energy analysis,
it might be adequate for considering the latter,
and then, we read the temporal balance equation of
$t_{\rm acc}={\rm min}(t_{\rm ad},t_{\rm syn},t_{\rm col})$,
which yields an enhanced value of $E_{t}$,
such that ${\rm min}(E_{s},E_{t})=E_{s}$ more likely appears.
The $E_{s}$-levels evaluated from Equation~(\ref{eq:7}), set as
$B=B_{{\rm eq},\delta=1}$, are indicated in Figure~4.
For convenience, we list the resulting $E_{p,m}$-values
in Table~\ref{tbl:2} (in round brackets).
In comparison with the results of $U_{\rm app}=0.1c$, the values
of $E_{p,m}$ increase, since the $E_{s}$-values are larger than
$E_{t}$ (determined from $E_{\rm sh}$), as seen in Figure~4.

The magnetic field strength involves the uncertainty,
coming about through some assumptions in the
estimation of the equipartition value (Section~2.2).
Thus we also investigate the field intensity
dependence of $E_{p/i,m}$.
In Figure~5, we show $E_{p,m}$ against $B_{{\rm eq},\delta=1}$
for $\theta=15\degr$; the linear relations simply stem from
$E_{s/t}\propto B(=B_{{\rm eq},\delta=1}\delta^{-5/7})$
in Equations~(\ref{eq:7}) and (\ref{eq:9}).
It is found that in the region with knot~BX5,
$E_{p,m}(E_{i,m})\geq 10^{20}~{\rm eV}$ can be achieved for the
ranges of $B_{{\rm eq},\delta=1}\geq 230~{\rm\mu G}(9~{\rm\mu G})$
and $\geq 85~{\rm\mu G}(3~{\rm\mu G})$ for $U_{\rm app}=0.1c$
and stationary shock cases, respectively.

\begin{figure}
\centerline{\includegraphics*[bb=120.0 32.5 630.0 590.0,
width=\columnwidth]{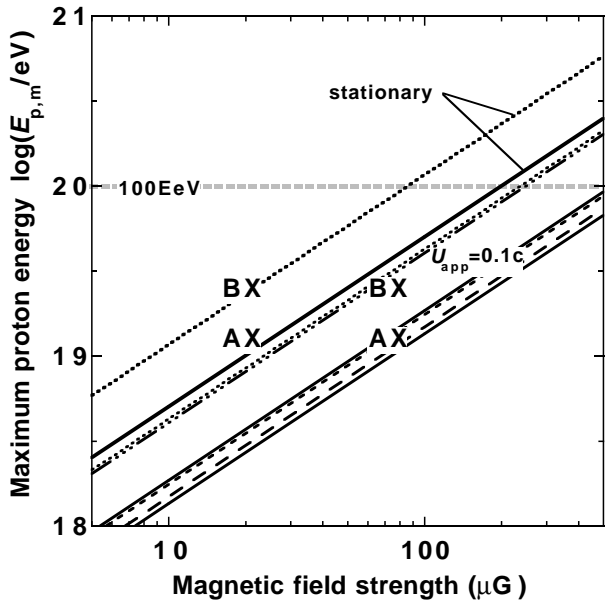}}
\caption{Maximum proton energy $E_{p,m}$ vs. the
magnetic field strength compared to an equipartition value
$B_{{\rm eq},\delta=1}(=B\delta^{5/7})$,
for $U_{\rm app}=0.1c$ and stationary shock cases.
The explanatory notes for the former are the
same as those given in Figures~2 and 3.\\
}
\end{figure}

\subsection{\it High-energy Neutrino Emissions}

The $p\gamma$ interaction discussed in Section~3.2.2 is revisited
in light of the major neutrino production mechanism.
The reaction sets in when the center of mass energy
of a $p\gamma$ interaction exceeds the threshold energy
for the $\Delta$-resonance; that is, $E_{p,m}$ is
required for exceeding the quantity of
$(m_{\Delta}^{2}-m_{p}^{2})c^{4}/4E_{\gamma}$ ($\equiv E_{p,{\rm th}}$),
where $m_{\Delta}$ and $E_{\gamma}$ are the mass of
the $\Delta$-particle and target photon energy,
respectively \citep[e.g.,][]{halzen}.
The threshold energy is found to scale as
$E_{p,{\rm th}}=1.6\times 10^{17}\Gamma^{2}
(1~{\rm eV}/E_{\gamma})~{\rm eV}$.
Above $E_{p,{\rm th}}$, charged and neutral pions can be
created by $p\gamma\rightarrow\Delta\rightarrow n\pi^{+}$
and $p\gamma\rightarrow\Delta\rightarrow p\pi^{0}$,
and the former triggers the decay chain of
$\pi^{+}\rightarrow\mu^{+}+\nu_{\mu}
\rightarrow e^{+}+\nu_{e}+{\bar\nu}_{\mu}+\nu_{\mu}$,
to produce the very high energy neutrinos.

Provided that the four final state leptons in the decay chain equally share
the pion energy, the threshold neutrino energy can be estimated as
$E_{\nu,{\rm th}}=\frac{1}{4}
\left< x_{p\rightarrow\pi}\right> E_{p,{\rm th}}$,
where $\left< x_{p\rightarrow\pi}\right>$
is the average fraction of energy transferred from the
initial proton to the produced pion \citep{halzen}.
By virtue of the active galactic nucleus (AGN) unification scheme
\citep[cf. Section~5;][]{urry,tsvetanov},
it might be justified to adopt the estimation of
$\left< x_{p\rightarrow\pi}\right>\simeq 0.2$,
which is adequate for blazar jet environments.
Then, for the target photon energy in the range of
$E_{\gamma}\sim 1~{\rm eV}$
\citep[for the Cen\,A jet; e.g.,][]{hardcastle06},
we work out at $E_{\nu,{\rm th}}\simeq 8\times 10^{15}~{\rm eV}$.
Similarly, one can estimate the achievable maximum
neutrino energy (denoted as $E_{\nu,m}$) with
reference to the $E_{p,m}$ values obtained above.
For convenience, the resultant $E_{\nu,m}$ values are
listed in Table~\ref{tbl:2}.
We find that $E_{\nu,m}$ is likely to be of EeV range in the
inner region; for $U_{\rm app}=0.1c$ and the stationary case,
the uppermost value is expected to reach as high as
$6\times 10^{18}~{\rm eV}$ and $2\times 10^{19}~{\rm eV}$,
respectively, in the region including knot~BX5.

Concerning the observability of the neutrino flux ($\phi_{\nu}$)
by a $\mu$-scintillation counter with the conversion probability
($P_{\nu\rightarrow\mu}$), the number of detected events could be
expressed as a function of neutrino energy, such that
$N_{\rm events}\sim\phi_{\nu}P_{\nu\rightarrow\mu}
\propto E_{\nu}^{-1/2}$ \citep{halzen}.
Since we have little knowledge of the actual efficiency
of energy conversion from accelerated protons to neutrinos
in FR-I radio jets, the spectral normalization of
$\phi_{\nu}$ involves the large uncertainty.
Considering the Auger flux, however, \citet{cuoco} attempted to calculate
the neutrino event rate in the detectors like IceCube, and found a rate
of $\sim 0.4-0.6~{\rm yr^{-1}}$ above a threshold of $10^{14}~{\rm eV}$.
In any case, the estimated $E_{\nu,m}$ provides the truncation
energy of the neutrino spectrum, and in turn, could influence
the neutrino flux estimation when the spectral index is given.
The related issue is important for clarifying the share of
cosmic-ray energy above the "knee" among celestial sources,
but seems to be somewhat out of scope in this paper.

\section{ON THE PARTICLE ACCELERATION IN THE LARGE-SCALE JET}

\subsection{\it Spatiotemporal Limits}

It may be worth applying the present DSA scenario to the
unsolved issue of particle acceleration in the
outer region more distant from the galactic core.
Here, we preliminarily show the $E_{\rm max}$
distribution in the large-scale jet, simply providing
$B$ as a parameter (without reference to the values
given in Table~\ref{tbl:1}), and spatial profile of $n$.
This might be more meaningful, if the length and radial scales
of jet preferentially determine $E_{\rm max}$, as we have seen
this property in the inner region (Section~3).
In the following, we investigate the condition, for which radiative
and collisional losses are negligible, to complete the discussions
on the feasibility of UHECR production in the large-scale jet
up to $\theta_{\rm proj}\simeq 200\arcsec$
\citep[$L_{\rm proj}\simeq 3.4~{\rm kpc}$, including the
outer knots C, E, F, and G; e.g.,][]{kraft02,hardcastle07}.

For the analysis in Section~4, we adopt the parameter values
listed in Table~\ref{tbl:3}, with the use of the
standard parameter set of $U_{\rm app}=0.1c$ and $p=2.4$.
We assume $\beta=5/3$ in the inner region including knots~A and B,
and $\beta=2$ in the outer region including knots~C, E, F, and G,
according to the arguments of the filamentary turbulent states
(Section~2.1): in Equations~(\ref{eq:7}) and (\ref{eq:9}),
for instance, the latter case $(\beta,p)=(2,2.4)$, which
reflects $(\alpha_{r},\alpha_{\rm x})=(0.7,1.2)$, gives
$a_{1}=2.4$, $a_{2}=3.8$, and ${\tilde\tau}=4.3\times 10^{-3}$.

\begin{table}
\begin{center}
\caption{Summary of the Parameter and $E_{\rm max}$ Values\label{tbl:3}}
\begin{tabular}{lccccl}
\tableline\tableline
Region & $L({\rm kpc})$\tablenotemark{a} &
$R({\rm pc})$\tablenotemark{b} & $\beta$ &
$E_{p,m}({\rm EeV})$\tablenotemark{c} &
$E_{i,m}({\rm ZeV})$\tablenotemark{c}\\
\tableline
A & $0.910$ & $352$ & $5/3$ & $6.5\,(25)$ & $0.17\,(0.66)$\\
B & $3.94$ & $819$ & $5/3$ & $21\,(59)$ & $0.54\,(1.5)$\\
C & $4.93$ & $1310$ & $2$ & $33\,(100)$ & $0.85\,(2.7)$\\
E & $7.29$ & $1310$ & $2$ & $40\,(100)$ & $1.0\,(2.7)$\\
F & $8.95$ & $2120$ & $2$ & $56\,(170)$ & $1.5\,(4.4)$\\
G & $13.4$ & $2120$ & $2$ & $69\,(170)$ & $1.8\,(4.4)$\\
\tableline
\end{tabular}
\tablenotetext{a}{The deprojected length, assuming the viewing angle of
$\theta=15\degr$ \citep{kraft02,hardcastle03}.}
\tablenotetext{b}{cf. \citet{kraft02}.}
\tablenotetext{c}{Explanatory note for the round brackets is the
same as that in Table~\ref{tbl:2}.}
\end{center}
\end{table}

In Figure~6, we plot $t_{\rm sh}$ and $t_{\rm syn}$, that is,
Equation~(\ref{eq:5}) for given $\theta$ as a parameter, and
Equation~(\ref{eq:6}) with $(A,Z)=(1,1)$ and $(56,26)$, given
$B_{{\rm eq},\delta=1}$ as a parameter (while $\theta=15\degr$ fixed).
It is clearly seen that as $L_{\rm proj}$ increases, the
value of $t_{\rm sh}$ monotonically increases; especially for
$\theta=15\degr$, from $t_{\rm sh}=3.2\times 10^{11}~{\rm s}$
(knot~A) to $4.7\times 10^{12}~{\rm s}$ (knot~G).
Note that the self-consistently determined $t_{\rm syn}$ level for iron
is comparable with that for proton, reflecting the physics that in the
original timescale (being proportional to $(A/Z)^{4}E^{-1}$), the
enhancement by the factor of $(A/Z)^{4}$ for iron is nearly canceled
by the increase of $E$, which is caused by both the weaker
synchrotron loss and stronger acceleration efficiency
(on account of the reduced $t_{\rm acc}$ ($\propto 1/Z^{2}$)).
Anyhow, it appears that, as far as the possible condition
of $B_{{\rm eq},\delta=1}\leq 500~{\rm\mu G}$
is retained, $t_{\rm sh}$ is smaller than $t_{\rm syn}$
in the entire region including knots A--G.

\begin{figure}
\centerline{\includegraphics*[bb=100.0 32.5 625.0 630.0,
width=\columnwidth]{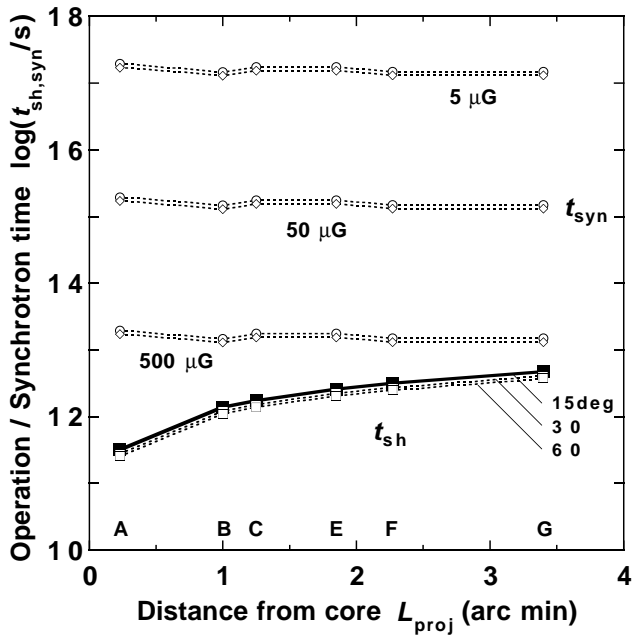}}
\caption{$t_{\rm sh}$ for given $\theta$ (labeled)
and $t_{\rm syn}$ for given $B_{{\rm eq},\delta=1}$ (labeled)
in each regions (labeled; up to the projected distance
of $\sim 200\arcsec$ from the core).
The open circles and diamonds indicate $t_{\rm syn}$
for proton $(A,Z)=(1,1)$ and iron $(56,26)$, respectively
(providing $\theta=15\degr$).
Note that $t_{\rm sh}<t_{\rm syn}$ is satisfied
for the conceivable ranges of $\theta$ and $B_{{\rm eq},\delta=1}$
throughout the concerned regions.\\
}
\end{figure}

As for the collisional loss, in parallel with the argument in
Section~3.2.3, we provide the critical proton density $n_{\rm cr}$,
above which $t_{\rm col}<t_{\rm sh}$ sets in.
For the timescale of nucleus--proton ($Np$) collision with
$t_{\rm col}\sim (\sigma_{Np}cn)^{-1}$,
we recall an empirical scaling of the cross section of
$\sigma_{Np}\sim\pi r_{0}^{2}A^{2/3}$, where
$r_{0}\approx 1.4\times 10^{-13}~{\rm cm}$.
In Figure~7, we explicitly depict $n_{\rm cr}$ against $L_{\rm proj}$
for $pp$ and $Np$ (say, $A=56$) collisions, for the case of $\theta=15\degr$.
The level of $n_{\rm cr}$ for $Np$ is lower than that for $pp$,
simply reflecting the thing that $t_{\rm col}$ for $Np$
is shorter than for $pp$ (by the factor of $\sim 0.04$).
The value of $n_{\rm cr}$ decreases as $L_{\rm proj}$
increases; e.g., for the $pp$ case, it ranges from
$n_{\rm cr}\simeq 3\times 10^{3}~{\rm cm^{-3}}$ (knot~A)
to $2\times 10^{2}~{\rm cm^{-3}}$ (knot~G).
Here, we attempt to make an ad hoc comparison
with a density distribution deduced from the beta model
that reasonably describes the galactic structure \citep{kraft02}.
By reasoning that the jet matter is concentrated on the
opening angle of $\phi$ ($\sim 2R/L$; i.e., the solid angle of
$\pi\phi^{2}/4$), we invoke the crude density profile (along
the jet) of $n(L_{\rm proj})\sim(16/\phi^{2})n_{b}(r)$,
where $L_{\rm proj}=r\sin\theta$ and
$n_{b}=n_{0}[1+(r/r_{G})^{2}]^{-3b/2}$ is
the beta-model density.
In Figure~7, we exemplify the density profile of $n=10^{2}n_{b}$,
reflecting $\theta=15\degr$ and $\phi\sim 20\degr$ (Figure~1).
Here, the parameter values of $n_{0}=4.0\times 10^{-2}~{\rm cm^{-3}}$,
$r_{G}=0.5~{\rm kpc}$, and $b=0.4$, have been adopted \citep{kraft02}.
We then find that $n$ takes the value in the range of
$\sim 10^{-1}-10^{0}~{\rm cm^{-3}}$ in between the region of knot~A to G
(in accord with a value suggested by \citet{meisenheimer}),
and the level is order of magnitude lower than the
$n_{\rm cr}$ level for both the $pp$ and $Np$ collisions.
Thus, it is claimed that the collisional effects on
the energy limit are negligible.
As a consequence, we can expect the significant reduction of
${\rm min}(t_{\rm sh},t_{\rm syn},t_{\rm col})\sim t_{\rm sh}$,
so that the temporal balance of $t_{\rm acc}=t_{\rm sh}$
provides the correct solution for $E_{t}$.

\begin{figure}
\centerline{\includegraphics*[bb=70.0 32.5 685.0 550.0,
width=\columnwidth]{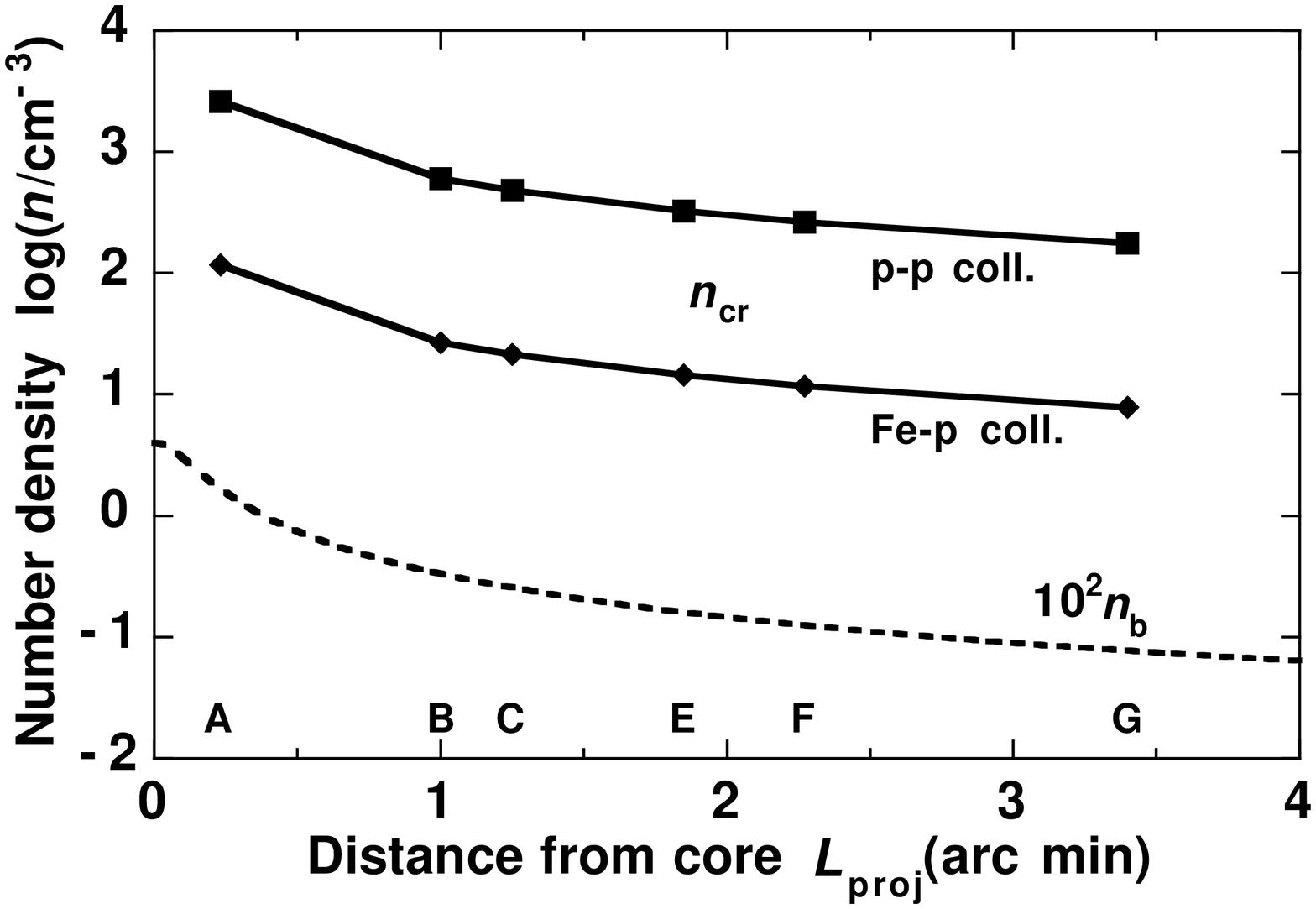}}
\caption{Critical density $n_{\rm cr}$ in each regions
(labeled), above which $t_{\rm col}<t_{\rm sh}$
(for $U_{\rm app}=0.1c$),
for the proton--proton and iron--proton collision cases (labeled).
For a comparison, the profile of the beta-model density $n_{b}$,
multiplied by factor $10^{2}$, is also depicted
(short dashed) as a function of $L_{\rm proj}$.
For further explanation, see the text.\\
}
\end{figure}

\subsection{\it Highest Energies}

For each regions in the large-scale jet, we evaluate the
achievable maximum energy of proton/iron, defined by
$E_{p/i,m}={\rm min}(E_{s},E_{t})$, where $E_{s}$ and $E_{t}$ are
given by Equations~(\ref{eq:7}) and (\ref{eq:9}), respectively.
In Figure~8, we plot $E_{s}$ and $E_{t}$ (for $Z=1$ and $26$),
providing $\theta=15\degr$ and
$B_{{\rm eq},\delta=1}=50~{\rm\mu G}$
\citep{romero,israel,hardcastle06,meisenheimer}
throughout the concerned regions.
The assumption of the $B_{{\rm eq},\delta=1}$ value, which
corresponds to a spatially averaged one, is adequate for
the present purpose, although the actual value might be
larger, particularly, in the inner jet region (Section~2.2).
One can see that ${\rm min}(E_{s},E_{t})=E_{t}$ is
satisfied as a whole, for both proton and iron.
The values of $E_{p/i,m}$ are summarized in Table~\ref{tbl:3}.
It seems more likely that, in the outer jet region,
$E_{p,m}$ can hardly reach $\sim 10^{20}~{\rm eV}$,
even though setting $R$-value to the uppermost range,
whereas $E_{i,m}$ exceeds $10^{20}~{\rm eV}$,
and reaches up to $2\times 10^{21}~{\rm eV}$.
When considering the stationary shock (in parallel to the arguments
in Section~3), the $E_{p/i,m}$ value increases, and particularly,
$E_{p,m}\sim 10^{20}~{\rm eV}$ is achieved in the outer region.
With regard to the neutrino production via the photopionization
decay chains, we read off
$E_{\nu,m}=\frac{1}{4}\left< x_{p\rightarrow\pi}\right>
E_{p,m}\gtrsim 10^{18}~{\rm eV}$ in the outer region.
Although the obtained results are preliminary, involving parameter
dependency, the expected overall features can be reconciled with the
manifestation (given in Sections~3.4 and 3.5) that the Cen\,A jet is a
candidate for the UHE proton accelerator that serves as EeV
neutrino factory, and simultaneously, for the iron Zevatron.

\begin{figure}
\centerline{\includegraphics*[bb=70.0 32.5 685.0 550.0,
width=\columnwidth]{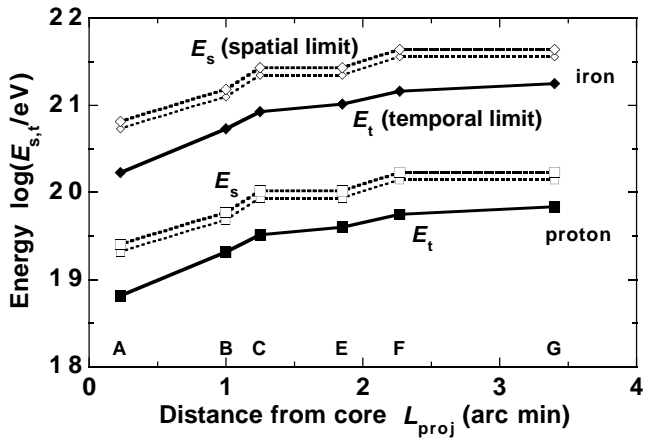}}
\caption{Energies $E_{s}$ (open marks) and $E_{t}$
(filled marks) determined by the spatial and temporal limits,
respectively, for proton and iron acceleration (labeled).
Here, $\theta=15\degr$ and $B_{{\rm eq},\delta=1}=50~{\rm\mu G}$
have been assumed throughout the concerned regions (labeled).
The filled marks indicate the maximum energy
obtained as $E_{\rm max}={\rm min}(E_{s},E_{t})=E_{t}$
for $U_{\rm app}=0.1c$, where $E_{s}$ is indicated by
the smaller marks; and the (larger) open ones indicate
$E_{\rm max}=E_{s}$ for the stationary shock case.
The $E_{\rm max}$ values (denoted as $E_{m,p}$ and $E_{m,i}$
for proton and iron, respectively)
are summarized in Table~\ref{tbl:3}.\\
}
\end{figure}

\section{DISCUSSION AND CONCLUSIONS}

The validity of the filamentary model has been checked by
reproducing the complicated blazar variabilities that
involve the non-monotonous hysteresis patterns of broadband
electromagnetic spectrum in flare phases \citep{honda08}.
By recalling the idea that regards FR-I radio sources
as misaligned BL\,Lac objects \citep{urry,tsvetanov},
the application of the filamentary model to
the Cen\,A jet will be reasonably justified.
As for the particle acceleration in the filamentary medium,
we point out that the present DSA mechanism for nuclei, which
owes to the (off-resonant) three-dimensional rms diffusion
across the shock (Section~3.1), is simpler than the electron
acceleration mechanism that is incorporated with the complexity of
the energy hierarchy mediated by the transition injection \citep{hh07}.
The acceleration of highest energy particle in the Cen\,A
was first considered by \citet{romero}.
In the early work, they dealt with the conventional gyro-resonant
diffusion model even for proton, that is to say, implicitly
supposed a large-scale ordered magnetic field
(with small-amplitude perturbations).
They addressed that a proton could be accelerated to the energy of
$2.7\times 10^{21}~{\rm eV}$ in situ, which is an order of magnitude
larger than the present $E_{p,m}$ values that are in the range of
$\sim 10^{19}-10^{20}~{\rm eV}$ (Section~3.4; Table~\ref{tbl:2}).
Such a larger value was derived from simply equating a classical
acceleration timescale \citep[e.g.,][]{biermann} with radiative
loss timescales, but this fashion now appears to be somewhat
optimistic (Section~3.2.2; Figure~2).
It can be claimed that taking into account of the shock operation
time or particle escape is essential for the maximum energy analysis
for the concerned source, to yield the reduced proton energy,
which can hardly reach the $10^{21}~{\rm eV}$ energy range.
It is also mentioned that the situation in which
the dynamical timescale limits the particle acceleration
is analogous to the situation that typically appears in
the supernova remnant shock acceleration, in which the
adiabatic expansion of spherical shell likely becomes
a major loss mechanism \citep[e.g.,][]{kobayakawa}.

In conclusion, we have evaluated the achievable maximum energies
of the proton and iron nucleus diffusively accelerated by
the shock in the Cen\,A jet including X-ray knots.
In particular, we have taken into account of the more realistic DSA
scenario that relies on the filamentary jet model responsible
for the recent X-ray measurements, and elaborated the
conceivable energy restriction stemming from spatiotemporal scales.
The key finding is that, for the plausible ranges of the
shock speed of $\sim 0.1c$, viewing angle of $\gtrsim 10\degr$,
and magnetic intensity of $\lesssim 500~{\rm\mu G}$,
the uppermost particle energy tends to be inevitably limited by
the shock accelerator operation timescale ($<10^{13}~{\rm s}$),
rather than the radiative cooling losses ($\gtrsim 10^{13}~{\rm s}$).
As a result, it has been demonstrated that there exists the
acceleration region (off the galactic core), in which proton
and iron can be energized to $\sim 10^{20}~{\rm eV}$
and $\sim 10^{21}~{\rm eV}$ ranges, respectively.
In particular, for the inner region including the X-ray knot BX5,
we work out at the maximum energies of $1\times 10^{20}~{\rm eV}$
and $3\times 10^{21}~{\rm eV}$ for proton and iron, respectively;
and estimate the corresponding cutoff energy of the neutrino produced
via the photopionization, as $6\times 10^{18}~{\rm eV}$.
For the stationary shock case, we can expect the enhancement of
the maximum energies, and read that the large-scale region
up to the projection of about $200\arcsec$, as well, has a potential
to energize proton and iron beyond $10^{20}~{\rm eV}$ and
$10^{21}~{\rm eV}$, respectively, to yield the $p\gamma$ neutrino
with the energy exceeding $5\times 10^{18}~{\rm eV}$.
We here manifest that the jet wide of the Cen\,A galaxy is a
promising candidate for the UHE proton accelerator and Zevatron
for high-$Z$ nucleus, as desirable to account for the outcome
from the Auger observatory \citep{abraham07}.

The derived large values of particle energies are
of those achievable at the acceleration sites.
The observable energies ought to be, of course, reduced,
due to a major energy loss mechanism including the
photopion interaction with cosmic microwave background.
However, the Cen\,A is so nearby that the ballistic
transport is anticipated to be not crucially degraded,
on account of the weak decay property
(such as $-(1/E)(dE/dt)\sim 10^{-8}~{\rm yr^{-1}}$
for $E>10^{20}~{\rm eV}$; \citet{romero}).
We hope, in near future, the application of the present scenario
to cosmologically distant (super GZK) AGNs, to solidify the point
source scenario for dominant UHECR production, which is responsible for
the suppression of cosmic-ray flux above $4\times 10^{19}~{\rm eV}$
that has recently been confirmed by the Auger's experiments
\citep{abraham08}, and also, is of interest in conjunction with
the detection of super GZK neutrinos \citep[e.g.,][]{ringwald}.

I am grateful to Y.~S.~Honda for a useful discussion.\\

\clearpage

\clearpage

\end{document}